\documentclass[twocolumn]{article}

\usepackage{graphicx}
\usepackage{physics}
\usepackage{siunitx}
\usepackage{graphicx}
\usepackage{comment}
\usepackage{subcaption}
\usepackage[normalem]{ulem}
\usepackage[cal=boondox]{mathalpha}
\usepackage{multirow}
\usepackage{booktabs}
\date{}

\author{
Dixith Manchaiah\thanks{dixith.manchaiah@nist.gov} \\
National Institute of Standards and Technology, Boulder, CO 80305, USA \\
Department of Physics, University of Colorado, Boulder, CO 80309, USA
\and
William J. Watterson \\
National Institute of Standards and Technology, Boulder, CO 80305, USA
\and
Christopher L. Holloway\thanks{christopher.holloway@nist.gov} \\
National Institute of Standards and Technology, Boulder, CO 80305, USA
}

\begin{document}
%\title{Frequency Comb Behavior from RF Driven Rydberg Dissipative Time Crystals}
\title{Frequency Comb Behavior of Time Crystals in an RF-Driven Dissipative Rydberg System}
\maketitle

%\author{Dixith Manchaiah}
%\email[]{dixith.manchaiah@nist.gov}
%\affiliation{National Institute of Standards and Technology, Boulder, Colorado 80305, USA}
%\affiliation{Department of Physics, University of Colorado, Boulder, Colorado 80309, USA}

%\author{William J. Watterson}
%\affiliation{National Institute of Standards and Technology, Boulder, Colorado 80305, USA}

%\author{Christopher L. Holloway}
%\email[]{christopher.holloway@nist.gov}

%\affiliation{National Institute of Standards and Technology, Boulder, Colorado 80305, USA}

\begin{abstract}
Driven nonlinear oscillators constitute a universal paradigm for understanding synchronization, frequency pulling, and frequency comb formation in nonequilibrium systems. Here, we realize such an emergent nonlinear oscillator in strongly interacting cesium Rydberg vapor, where coherent optical excitation, dissipation, and long-range interactions give rise to a driven-dissipative time crystal phase with intrinsic oscillation frequencies. Applying a radio-frequency (RF) field allows controlled tuning of the intrinsic oscillation frequency. Under RF heterodyne conditions, we observe intermodulation, frequency pulling, and, at strong drive, the emergence of a comb-like spectrum in the atomic coherence. We quantitatively capture these observations using a four-level mean-field model and demonstrate a classical analogue with a driven Van der Pol oscillator. Our results establish interacting Rydberg ensembles as a tunable platform for exploring nonequilibrium time-crystalline order, nonlinear synchronization, and frequency comb generation in many-body atomic systems.
\end{abstract}

\section{Introduction}

Unique states of matter are ubiquitous in nature. Conventional crystals arise from the spontaneous breaking of continuous spatial translation symmetry, leading to long-range order in space. In contrast, a distinct phase can emerge when time-translation symmetry is spontaneously broken. This state is known as a time crystal~\cite{Wilczek:2012jt,PhysRevLett.109.160402,Sacha_2018}. Such behavior has been observed across a variety of platforms, including driven many-body systems operating far from equilibrium, where collective dynamics play a crucial role~\cite{2017Natur.543..221C,Zhang:2016kpq,Ho:2017nee,Dreon:2021jia}. In these systems, competition between driving and dissipation can give rise to stable oscillatory behavior, which is imprinted not by an external drive but instead arises intrinsically from the system’s dynamics~\cite{Kessler:2021eol,Zhu:2019kny,Wadenpfuhl:2023ywo}. These self-oscillating systems exhibit robust nonlinear properties including multistable states, bifurcations, and related dynamical phenomena and can be described within the mathematical framework of nonlinear dynamics and bifurcation theory~\cite{https://doi.org/10.1155/2020/8510930}. %\sout{In particular, Hopf bifurcations explain the transition from a fixed point steady-state attractor to a stable limit cycle regime associated with the onset of persistent oscillations and time crystal behavior.} 

A natural extension concerns ensembles of self-sustained oscillators with an external periodic driving. In this case, the interplay between the intrinsic oscillatory frequency and the external driving frequency can lead to frequency and phase-locking, resulting in collective synchronization. In periodically driven many-body systems, such synchronization and frequency locking can manifest as discrete time-translation symmetry breaking. From a classical perspective, such collective phase locking can be captured by the Kuramoto model~\cite{STROGATZ20001}, which describes the emergence of collective order in ensembles of globally coupled oscillators with distributed natural frequencies~\cite{RevModPhys.77.137}. 

Rydberg atomic ensembles provide a platform to explore such driven, many-body dynamics due to their strong interactions and tunable dissipation~\cite{Wadenpfuhl:2023ywo}. Their versatility has enabled the exploration of non-equilibrium phase transitions and critical behavior in interacting quantum systems~\cite{PhysRevA.94.011401,Glaetzle:2012bbb,Wu:2023evo}. Recent studies of driven-dissipative Rydberg systems and related open quantum platforms have revealed a rich landscape of nonequilibrium phenomena, including non-equilibrium phase transitions~\cite{PhysRevLett.111.113901}, bistability~\cite{PhysRevA.94.011401}, self-organized dynamics~\cite{PhysRevX.10.021023}, and ergodicity breaking~\cite{Ding:2023kbb}. Within this framework, collective synchronization and frequency locking have been shown to give rise to time crystal behavior, with reports of bifurcations of time crystal phases~\cite{Liu:2024kfa} and the coexistence of multiple time crystal orders~\cite{Jiao:2024cvh}. In parallel, realizations such as electric-field sensing~\cite{Arumugam_2025} and Stark-modulated time crystals for sub-kHz detection~\cite{Arumugam_2025stark} have been demonstrated.

\begin{figure*}[h]%
		\centering
		\includegraphics[width=\linewidth]{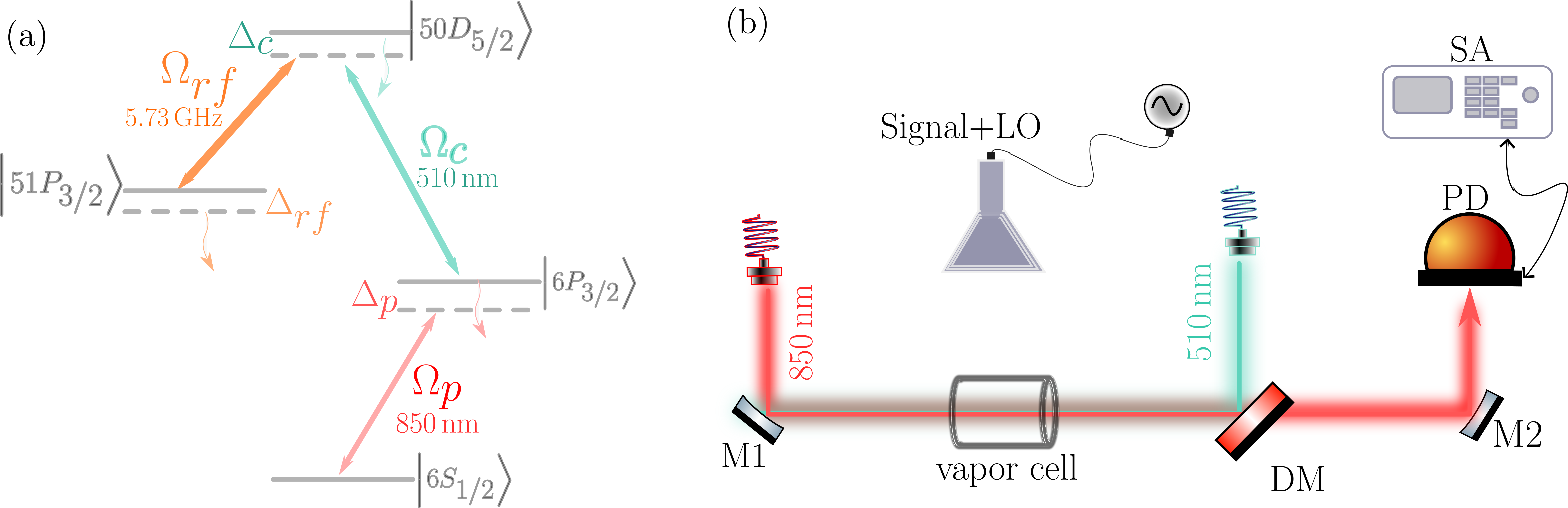}
		\caption{(a) Energy-level diagram of the four-level cesium system driven by optical and RF fields. (b) Experimental setup for two-photon excitation of cesium atoms. The probe laser is shown in red and coupling laser is shown in teal, counter-propagates with respect to each other. The rf signal and LO fields are applied using a horn antenna. M1, M2: mirrors; DM: dichroic mirror; PD: photodetector; SA: spectrum analyzer; Signal+LO: combined rf signal and LO fields.}
		\label{fig:b}
\end{figure*}

In this work, we investigate a driven-dissipative time crystal phase in a cesium Rydberg vapor system under radio-frequency (RF) driving, where strong interactions, dissipation, and external modulation collectively shape the nonequilibrium dynamics. By applying an RF field to the Rydberg transition, we demonstrate that the intrinsic oscillation frequency of the time crystal response can be continuously tuned via Stark modulation. Under RF heterodyne conditions, we observe frequency pulling and the emergence of a frequency comb structure in the collective optical response of the ensemble indicating the presence of multiple phase-locked temporal modes. The appearance of these regularly spaced spectral components reflects nonlinear synchronization and frequency locking in the interacting many-body system. To elucidate the underlying mechanisms, we develop a theoretical description that captures the time-crystal phase and interpret the experimental observations within a driven nonlinear classical oscillator framework. % This approach establishes clear connections between the observed comb formation and classical concepts such as limit-cycle dynamics and synchronization. 
 While time-crystal behavior and synchronization have been explored in a variety of driven quantum platforms, their manifestation in warm Rydberg vapor ensembles accessed through RF heterodyne detection has remained largely unexplored. In particular, the emergence of narrow-band frequency comb structures arising from synchronization-driven nonlinear collective dynamics has not been previously reported in this setting. Our results demonstrate a tunable driven-dissipative time crystal in a cesium vapor system and gives rise to frequency-comb behavior, opening new avenues for exploring nonequilibrium many-body dynamics in Rydberg media and low-frequency RF sensing.

\section{Methods}

We employ a two-photon optical excitation scheme to drive cesium atoms to a Rydberg state, as illustrated in Fig.~\ref{fig:b}a. The experimental setup is shown in Fig.~\ref{fig:b}b. A probe laser at \SI{850}{\nano\meter} and a coupling laser at \SI{510}{\nano\meter} are derived from external-cavity diode lasers (ECDLs). The coupling laser is generated via second-harmonic generation in a cavity from a fundamental ECDL operating at \SI{1020}{\nano\meter}. The probe laser is frequency locked to the saturated absorption spectrum of the cesium D$2$ transition, addressing the $\ket{6S_{1/2},F=4} \rightarrow \ket{6P_{3/2},F'=4,5}$ hyperfine transitions. The coupling laser at \SI{510}{\nano\meter} is used to drive the $\ket{6P_{3/2}} \rightarrow \ket{50D_{5/2}}$ Rydberg transition. A \SI{7}{\centi\meter}-long cesium vapor cell is placed at the intersection of the probe and coupling beams. The two beams counter-propagate to reduce Doppler broadening. The probe beam is focused to a $1/e^2$ waist of \SI{150}{\micro\meter} and has a fixed Rabi frequency of \SI{150}{\mega\hertz} throughout the experiment. The coupling beam has a $1/e^2$ waist of \SI{200}{\micro\meter}, and its power is varied to explore different excitation regimes. Care is taken to ensure that the probe and coupling beams are well overlapped throughout the interaction region inside the vapor cell. After interacting with the atoms, the probe beam passes through a dichroic mirror and is collected using a photodetector (PD). The coupling laser is scanned across the Rydberg transition while monitoring the probe transmission on the PD. Under two-photon resonance conditions, an electromagnetically induced transparency (EIT) peak corresponding to the $\ket{50D_{5/2}}$ state is observed, along with an additional peak associated with the $\ket{50D_{3/2}}$ state. These peaks are used to calibrate the frequency axis. The vapor cell temperature is maintained at \SI{50}{\degreeCelsius}. The RF field is applied to the vapor cell by generating one or two tones from a signal generator, sending the tones through a power combiner, and transmitting the signal from a horn antenna placed 30~cm from the vapor cell.

\section{Results}
\subsection{Nonlinear response under coupling field excitation}
\begin{comment}
\begin{figure}[t]
	\centering
		\includegraphics[width=\linewidth]{tc.png}
		\caption{Probe transmission as a function of the coupling detuning for different coupling Rabi frequencies.}
		\label{fig:tc}
\end{figure}
	
	\begin{figure}[t]
		\centering
		\includegraphics[width=\linewidth]{oscillations}
		\caption{Time domain probe transmission for different coupling laser detunings, showing the emergence of self-sustained oscillations.}
		\label{fig:osc}
        \end{figure}
\end{comment}

    \begin{figure*}[t]
		\centering
		\includegraphics[width=\linewidth]{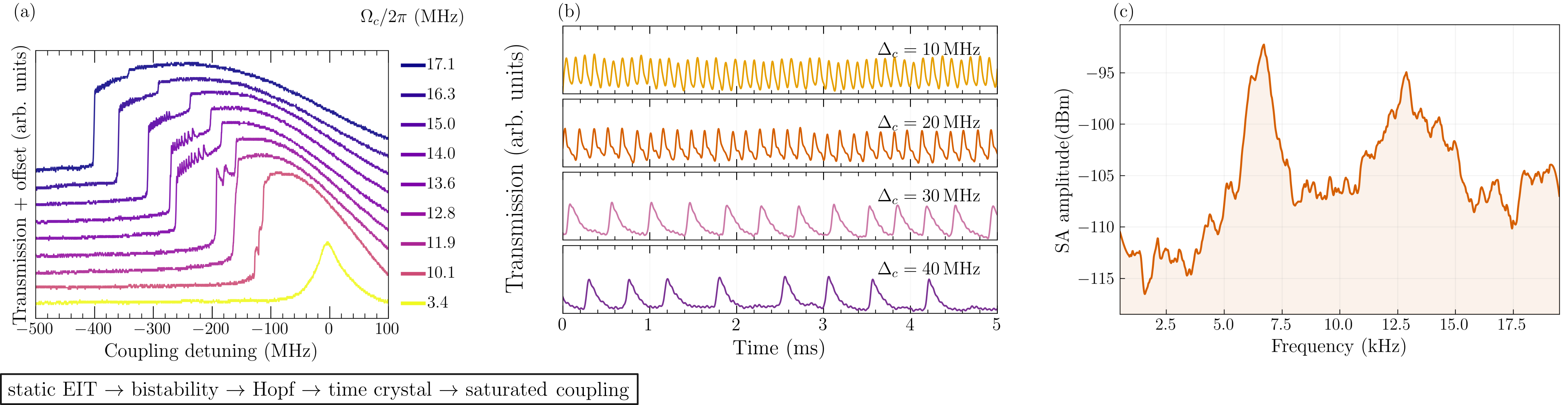}
		\caption{(a) Probe transmission as a function of the coupling detuning for different coupling Rabi frequencies. (b) Time domain probe transmission for different coupling laser detunings, showing the emergence of self-sustained oscillations. (c) Frequency domain spectrum of the time domain signal corresponding to $\Delta_c$ = 10 MHz shown in Fig. \ref{fig:1}b  measured using a spectrum analyzer.}
		\label{fig:1}
	\end{figure*}
Figure \ref{fig:1}a shows the probe transmission spectrum as a function of coupling detuning for different coupling Rabi frequencies without any applied RF. At a low coupling Rabi frequency of approximately $\SI{3.4}{\mega\hertz}$ (shown in yellow), a single Lorentzian-like transparency peak is observed. This feature can be attributed to a power-broadened EIT response, arising from the relatively strong probe field compared to the coupling field. As the coupling Rabi frequency is increased to $\SI{10.1}{\mega\hertz}$, the EIT feature becomes broader and develops an asymmetric lineshape. In this regime, the population in the Rydberg state increases and interactions between Rydberg atoms begin to influence the optical response, placing the system in a nonlinear EIT regime. At a coupling Rabi frequency of $\SI{11.9}{\mega\hertz}$, the system exhibits a clear transition marked by an abrupt vertical jump in transmission near $\Delta_c = \SI{-160}{\mega\hertz}$. This behavior indicates the onset of bistability, where the increase in Rydberg population enhances Rydberg-Rydberg interactions and provides strong feedback to the effective detuning of the system. For coupling Rabi frequencies between  $\SI{12.8}{\mega\hertz}$ and $\SI{15.0}{\mega\hertz}$, ripples appear in the transmission spectrum. In this regime, the system shows clear oscillatory behavior and evolves into a stable self-sustained oscillations, corresponding to the emergence of a time-dependent dynamical state. As shown in the Fig.\ref{fig:1}a, a wide plateau region develops, indicating strong nonlinear feedback from Rydberg-Rydberg interactions and the formation of a dissipative time crystal like phase. Further increasing the coupling Rabi frequency leads to a widening of the plateau region and a gradual smoothing of the oscillatory features, consistent with the system entering a strongly coupled, saturated time crystal regime.
The observed transmission spectra reflects the increasing role of interaction-induced nonlinear feedback in the system. As the coupling strength is increased, the growing Rydberg population leads to a population-dependent shift of the Rydberg level, which modifies the effective two-photon detuning and gives rise to asymmetric line shapes and bistable \cite{PhysRevLett.111.113901}. In an intermediate coupling regime, this feedback destabilizes the steady-state solution, resulting in self-sustained oscillations and the emergence of a time crystal like phase. At higher coupling strengths, the nonlinear response becomes dominant over a wider detuning range, leading to the formation of extended plateau regions and a saturated oscillatory phase.

\subsection{Detuning dependent oscillatory dynamics}
At a coupling Rabi frequency of $\SI{14.0}{\mega\hertz}$ and a detuning of $\SI{-220}{\mega\hertz}$, the coupling laser is frequency-locked using a wavemeter-based PID feedback for frequency stability. The time-domain probe transmission measured at fixed coupling detunings is shown in Fig.~\ref{fig:1}b. At a detuning of $\Delta_c=\SI{10}{\mega\hertz}$ from the frequency locked region, small-amplitude and nearly sinusoidal oscillations with a constant amplitude are observed, indicating the onset of self-sustained oscillations close to the instability threshold. Increasing the detuning to $\Delta_c=\SI{20}{\mega\hertz}$ results in oscillations with sharper peaks and asymmetric waveforms, suggesting an increased influence of nonlinear feedback associated with interaction-induced shifts of the Rydberg level. At $\Delta_c=\SI{30}{\mega\hertz}$, the oscillations become strongly anharmonic and exhibit relaxation-type dynamics, characterized by extended intervals of low transmission interrupted by sudden, large-amplitude spikes. For a larger detuning of $\Delta_c=\SI{40}{\mega\hertz}$, the dynamics consists of sparse, high-amplitude pulses separated by long quiescent periods.

The systematic evolution of the oscillation waveform with increasing detuning reflects the changing balance between optical excitation, dissipation, and interaction-induced feedback in the system. At smaller detunings, the steady state is only weakly unstable, giving rise to nearly harmonic oscillations. As the detuning increases, the growing interaction-induced shift of the Rydberg level enhances the nonlinear feedback, leading to waveform asymmetry and the emergence of relaxation type oscillations. At larger detunings, a clear separation of timescales develops, where slow population buildup is followed by rapid switching~\cite{Wadenpfuhl:2023ywo}. Together, these observations demonstrate that the system evolves into a robust limit-cycle regime, consistent with the stable self-sustained oscillatory behavior underlying the plateau and ripple features observed in the transmission spectra.

At this point, the coupling laser was frequency-locked at a detuning of $\Delta_c=\SI{10}{\mega\hertz}$, corresponding to a stable oscillatory regime, and the probe transmission detected on a photodiode was analyzed in the frequency domain using a spectrum analyzer. Figure~(\ref{fig:1}c) shows the resulting spectrum, which exhibits a dominant peak at approximately $\SI{6.5}{\kilo\hertz}$ along with a secondary, broader peak centered around $\SI{13}{\kilo\hertz}$. The observed oscillations depends on system parameters such as interaction strengths, number density and Rabi frequencies~\cite{Arumugam_2025}. The dominant peak corresponds to the fundamental frequency of the self-sustained oscillations observed in the time-domain measurements, while the broader secondary peak arises from the non-sinusoidal nature of the oscillations, which generates higher harmonics in the frequency domain due to the nonlinearity. The spectrum shown is obtained by averaging over five repeated measurements to improve the SNR. 
%
\begin{comment}

\begin{figure}[h]
	\centering
	\includegraphics[width=\linewidth]{saplot}
	\caption{Frequency domain spectrum of the time domain signal corresponding to $\Delta_c$ = 10 MHz shown in Fig. \ref{fig:osc}  measured using a spectrum analyzer.}
	\label{fig:saplot}
\end{figure}
\begin{figure}[h]
	\centering
	\includegraphics[width=0.9\linewidth]{RF1}
	\caption{Two dimensional color map of the RF power as a function of frequency, showing the evolution of the oscillation frequency with increasing RF drive strength.}
	\label{fig:rf1}
\end{figure}
\end{comment}
\subsection{RF driven frequency tuning of self-sustained oscillations}

\begin{figure*}[h]
	\centering
	\includegraphics[width=\linewidth]{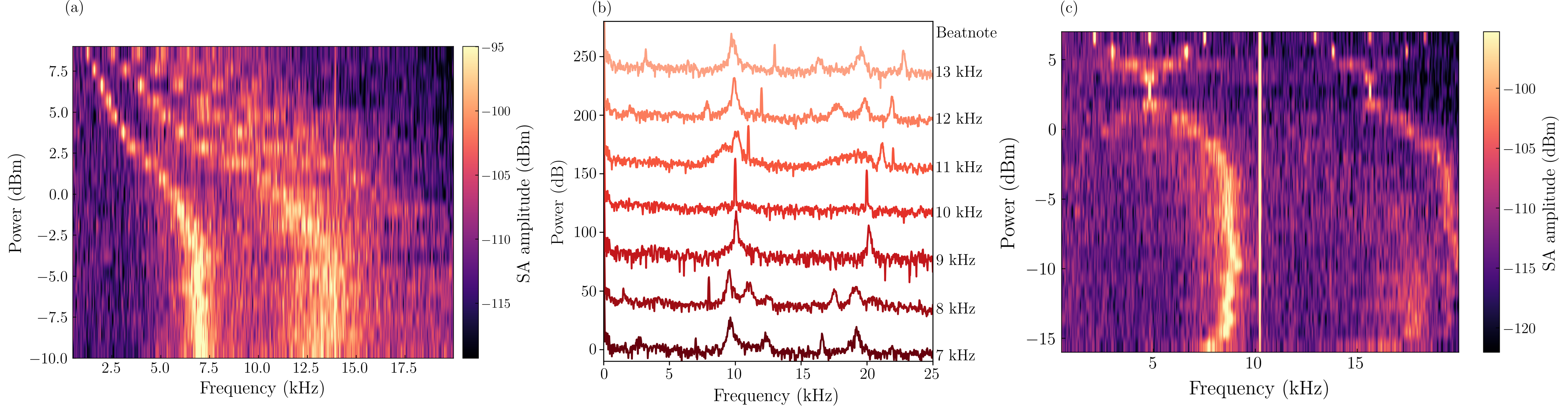}
	\caption{(a) Two dimensional color map of the RF power as a function of frequency, showing the evolution of the oscillation frequency with increasing RF drive strength. (b) Intermodulation products and injection locking of the stable oscillation phase mixing with the RF heterodyne beatnote. The stable oscillation phase has a main component near 10 kHz and a harmonic at 20 kHz. The primary beatnotes are indicated by the legend and the mixed products occur at the sum and difference frequencies between the beatnote and stable oscillations. The traces are offset vertically for clarity with a similar noise floor of -120 dBm for each trace. (c) Two dimensional color map measured under heterodyne conditions, showing the signal amplitude as a function of frequency and RF power. The LO power is fixed at 0 dBm, and detuned by 10 kHz from RF signal field.}
	\label{fig:heterodyne}
\end{figure*}

An external RF field at $\SI{5.73}{\giga\hertz}$ is applied to drive the $\ket{50D_{5/2}} \rightarrow \ket{51P_{3/2}}$ transition using a signal generator connected to a horn antenna. Since our focus is on the dynamics of the stable self-sustained oscillatory regime, the RF frequency is detuned by $\SI{280}{\mega\hertz}$. In this Stark-shift regime, the RF field is applied while its power is varied, and the resulting probe transmission spectrum in the frequency domain is shown as a 2D map in Fig.\ref{fig:heterodyne}a. At low RF power, a well defined dominant oscillation peak at approximately $\SI{6.5}{\kilo\hertz}$ is observed, corresponding to the intrinsic self-sustained oscillations of the system. This peak, along with its harmonics, remains largely unaffected by the applied RF power in this regime. As the RF power is increased, the intrinsic oscillation frequency gradually shifts toward lower frequencies. This shift is continuous over the explored RF power range and reflects the increasing influence of the RF-induced Stark shift on the energy levels. In addition to this frequency shift, new spectral features appear at higher frequencies as the RF power is increased, indicating a modification of the oscillatory dynamics by the applied field. These features correspond to higher harmonics or mixed frequency components arising from the nonlinear response of the atomic system. Furthermore, the main oscillation peak narrows with increasing RF power, as seen in the bright band of frequency. At the highest RF powers, narrow and more structured spectral features become clearly visible, indicating a frequency response of the driven oscillatory system.

The introduction of an RF field leads to an interplay between the intrinsic self-sustained oscillations and the RF-induced Stark shift and dressing of the Rydberg states. As the RF power is increased, the total shift modifies the effective detuning of the Rydberg transition, resulting in a gradual tuning of the intrinsic oscillation frequency toward lower values. The appearance of multiple oscillation frequencies at higher RF powers is a signature of the onset of nonlinear mixing between the intrinsic oscillations and the RF-induced modulation~\cite{Jiao:2024cvh}. These features become more pronounced and well defined at higher RF powers. The application of the RF field thus enables controlled tuning of the oscillation frequencies through modification of the effective detuning of the system.

\subsection{RF heterodyne control of self-sustained oscillations}
The appearance of multiple frequency components at higher RF power motivated us to explore the system further by introducing an additional local oscillator (LO) and signal (SIG) RF field. We first investigated the effect of varying SIG detuning at fixed LO power of 5 dBm and SIG power of 0 dBm (Fig. \ref{fig:heterodyne}b). When the frequency difference between the stable oscillatory frequency, $f_{osc}$, near 10 kHz and the beatnote frequency, $f_{bn}$ are greater than roughly 1kHz, we see intermodulation frequencies, $f_{IF}$, occurring at 

\begin{equation}
    f_{IF} = |nf_{osc} \pm f_{bn}|
    \label{harmonicmixer}
\end{equation}
\\
The above equation, (Eq. \ref{harmonicmixer}), describes an up and down-conversion harmonic mixer. At beatnotes of 9 kHz and 10 kHz, near the stable oscillatory frequency, we see injection locking of the oscillatory phase to the beatnote frequency. Similar observations were made by Arumugam for very low frequency (VLF) fields applied to the atoms through a capacitor plate internal to the glass vapor cell \cite{arumugam2025injection}.
\begin{comment}
    \begin{figure}[h]
	\centering
	\includegraphics[width=\linewidth]{figures/Intermodulation.png}
	\caption{Intermodulation products and injection locking of the stable oscillation phase mixing with the RF heterodyne beatnote. The stable oscillation phase has a main component near 10 kHz and a harmonic at 20 kHz. The primary beatnotes are indicated by the legend and the mixed products occur at the sum and difference frequencies between the beatnote and stable oscillations. The traces are offset vertically for clarity with a similar noise floor of -120 dBm for each trace.}
	\label{fig:Intermodulation}
\end{figure}

\begin{figure}[h]
	\centering
	\includegraphics[width=\linewidth]{heterodyne}
	\caption{Two dimensional color map measured under heterodyne conditions, showing the signal amplitude as a function of frequency and RF power. The LO power is fixed at 0 dBm, and detuned by 10 kHz from RF signal field.}
	\label{fig:heterodyne}
\end{figure}

\end{comment}

Having investigated the dynamics for fixed LO and SIG power with varying beatnote frequency, we next fixed the beatnote frequency and tuned the SIG power. The LO field is applied at a fixed power of $\SI{0}{dBm}$ and detuned by $\SI{10}{\kilo\hertz}$ from the applied SIG field. As shown in Fig.~\ref{fig:heterodyne}c, the $\SI{10}{\kilo\hertz}$ beatnote appears as a single narrow spectral feature and remains well defined over the entire range of applied signal RF powers. In addition to the beatnote, intrinsic self-sustained oscillations are observed at $\SI{7.5}{\kilo\hertz}$. 
At low SIG RF power, the amplitude of the intrinsic oscillation dominates over that of the beatnote. As the SIG power is increased, the intrinsic oscillation shifts toward the beatnote frequency, indicating a clear modification of the oscillatory dynamics by the applied RF field. With further increase in the SIG RF power, the intrinsic oscillation moves away from the beatnote and, around a SIG power of $\SI{0}{dBm}$, collapses into narrow and distinct spectral peaks at approximately $\SI{5}{\kilo\hertz}$, $\SI{10}{\kilo\hertz}$, and $\SI{15}{\kilo\hertz}$. The higher harmonics follow a similar evolution to that of the intrinsic oscillation peak. Interestingly, upon increasing the SIG power further, multiple well-defined and equally spaced frequency components emerge, forming a comb-like spectrum in the frequency domain. This comb structure occurs when $f_{bn}$ frequency is an integer multiple of $f_{osc}$, indicating the intermodulation products seeding multiple phase locked temporal modes in the atomic coherence.  %At even higher SIG powers, the comb features broaden and their contrast relative to the background is reduced, indicating a gradual disruption of the comb-like structure.
\begin{figure*}[h]
	\centering
	\includegraphics[width=\linewidth]{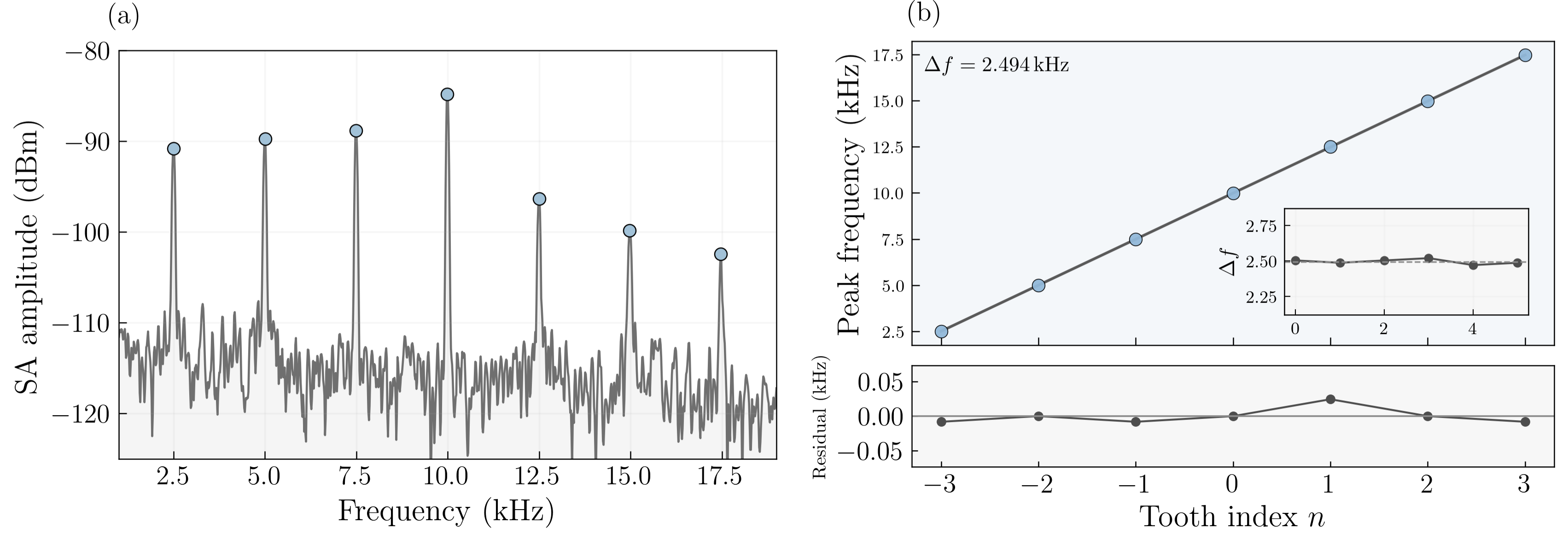}
	\caption{(a) Frequency comb-like structure in the heterodyne spectrum for a LO RF power of 0 dBm and a signal RF power of 5 dBm. (b) Peak frequency versus comb tooth index. The data points are fitted to the frequency comb equation (solid line). The inset shows the frequency spacing between adjacent teeth, and the bottom panel shows the residuals versus tooth index.}
	\label{fig:comb}
\end{figure*}
Figure~\ref{fig:heterodyne}c demonstrates that the simultaneous application of SIG and LO fields enhances the nonlinear interaction between the intrinsic self-sustained oscillations and the applied RF fields. In contrast to conventional heterodyne schemes, the LO field in this configuration primarily serves as a stable frequency reference rather than fixing the Rydberg level splitting, while the SIG field plays the dominant role in modifying the oscillatory dynamics. As the SIG RF power is varied, the RF-induced Stark shift alters the effective detuning of the Rydberg transition, which in turn influences the intrinsic oscillations and leads to frequency pulling toward the beatnote frequency. At low SIG RF powers, the nonlinear coupling is weak and the intrinsic oscillation remains the dominant spectral feature. As the SIG power is increased, the nonlinear interaction between the intrinsic oscillation and the heterodyne fields is enhanced while coherence is preserved, resulting in the emergence of additional, evenly spaced frequency components. These observations indicate that controlled variation of the SIG RF power enables access to a regime of strong nonlinear mixing, giving rise to a comb-like spectral structure in the frequency domain.
%
\begin{comment}
    \begin{figure}[h]
	\centering
	\includegraphics[width=0.9\linewidth]{comb}
	\caption{Frequency comb-like structure in the heterodyne spectrum for a LO RF power of 0 dBm and a signal RF power of 5 dBm.}
	\label{fig:comb}
\end{figure}

\end{comment}

Figure~\ref{fig:comb}a shows comb-like spectrum at the RF SIG power of $\SI{5}{dBm}$. Distinct well-resolved peaks are observed at discrete frequencies. The peaks are equally spaced in frequency with  a spacing of $\SI{2.5}{\kilo\hertz}$. The central peak at $\SI{10}{\kilo\hertz}$ is the heterodyne beatnote frequency and all the additional peaks appear symmetrically around it. To quantify the comb-like spectrum, the peak positions of discrete frequency spectrum are extracted and it is used to fit to the equation
\begin{equation}
	f_n = f_0 +n\Delta f.
\end{equation} 
The extracted peak frequencies follow a linear dependence on tooth index $n$. Fig.\ref{fig:comb}b shows the plot of peak frequency versus the tooth index. The linear fit yields a mean tooth spacing of $\Delta f =\SI{2.496}{\kilo\hertz}$. The inset plot shows the fluctuation of $\Delta f$ around $\SI{2.5}{\kilo\hertz}$. Residuals are shown and it remains near zero with small deviations, showing that the tooth frequencies are well described by uniform spacing.

The appearance of equally spaced spectral components reflects nonlinear frequency mixing between the intrinsic self-sustained oscillation and the heterodyne RF fields. In this regime, the system supports a stable oscillatory state that acts as a nonlinear mixer, generating sidebands around the heterodyne beatnote at integer multiples of the characteristic mixing frequency. The uniform spacing and small residuals indicate that the underlying oscillatory dynamics remain coherent over the measurement timescale, allowing well-defined frequency components to form. The observation of a linear dependence of the peak frequencies on the tooth index confirms that the comb-like spectrum arises from a deterministic nonlinear response rather than from broadband noise. In total, these results demonstrate controlled access to nonlinear oscillatory dynamics and comb-like spectral responses in a driven-dissipative Rydberg system.

\section{Discussion}
The results demonstrate that two-photon Rydberg excitation in an EIT system enables access to nonlinear dynamical regimes ranging from bistability to self-sustained oscillations and a time-crystal phase. External RF and heterodyne driving further allow controlled modification of the intrinsic oscillation frequency and the emergence of comb-like spectral features. The emergence of self-sustained oscillations in the present system originates from nonlinear feedback between the Rydberg population and the effective optical detuning in a driven-dissipative environment. Under strong optical excitation, interactions between Rydberg atoms induce a population-dependent shift of the Rydberg level, which in turn modifies the excitation rate. This feedback destabilizes the steady-state solution and gives rise to a stable limit-cycle oscillation. Dissipation through spontaneous decay and dephasing prevents runaway excitation and stabilizes the oscillatory dynamics, resulting in persistent intrinsic oscillations without external modulation. Such limit-cycle behavior is a characteristic feature of dissipative time-crystal phases and underlies the time-domain oscillations observed in the experiment~\cite{Wadenpfuhl:2023ywo,Arumugam_2025,Jiao:2024cvh,Liu:2024kfa}.

To support this interpretation, we consider a mean-field model based on a four-level ladder system with interaction-induced feedback incorporated through an effective Rydberg level shift. The Hamiltonian for the four-level ladder scheme for a given velocity class $v$ is given as~\cite{Wadenpfuhl:2023ywo, holloway2017electric},

%\begin{comment}
%\begin{equation}
	%H^{(v)}(t) = \frac{\hbar}{2}
	%\begin{pmatrix}
%		0 & \Omega_p & 0 & 0 \\[4pt]
%		\Omega_p & -2\Delta_p^{(v)} & \Omega_c & 0 \\[4pt]
%		0 &\Omega_c & -2\big(\Delta_p^{(v)} + \Delta_c^{(v)} -  V_3^{MF}\big) & \Omega_{RF}\\[4pt]
%        0 & 0 & \Omega_{RF} & -2\big(\Delta_p^{(v)} + \Delta_c^{(v)} + \Delta_{RF}%\end{pmatrix}
%\end{equation}
%\end{comment}

\begin{equation}
H^{(v)}(t) = \frac{\hbar}{2}
\begin{pmatrix}
0 & \Omega_p & 0 & 0 \\
\Omega_p & -2\Delta_p^{(v)} & \Omega_c & 0 \\
0 & \Omega_c &
-2\tilde{\Delta}_3^{(v)} & \Omega_{RF} \\
0 & 0 & \Omega_{RF} &
-2\tilde{\Delta}_4^{(v)}
\end{pmatrix}
\end{equation}
\begin{align}
\tilde{\Delta}_3^{(v)} &= \Delta_p^{(v)} + \Delta_c^{(v)} - V_3^\text{MF}, \\
\tilde{\Delta}_4^{(v)} &= \Delta_p^{(v)} + \Delta_c^{(v)}
+ \Delta_{RF}^{(v)} - V_3^\text{MF} - V_4^\text{MF}.
\end{align}
where the velocity dependent probe and coupling detunings are given by
\begin{equation}
	\Delta_p^{(v)} = \Delta_p - k_p v_j,
	\qquad
	\Delta_c^{(v)} = \Delta_c + k_c v_j.
\end{equation}
Here, $\Delta_p$ and $\Delta_c$ are the probe and coupling detunings at zero velocity, $\Omega_p$ and $\Omega_c$ are the probe and coupling Rabi frequencies, and $k_p$ and $k_c$ are the wave vectors corresponding to probe and coupling fields, respectively. Here $V_{i}^\text{MF}$ denotes the mean-field interaction induced energy shift of the Rydberg $i_\text{th}$ state due to Rydberg-Rydberg interactions and it is given as,
\begin{equation}
	V_{i}^\text{MF}=V_{\mathrm{norm}}\big(\langle\rho_{ii}(t)\rangle\big)^{\beta},
    \label{V_MF}
\end{equation}
where $\rho_{ii}$ denotes the Rydberg-state population, $V_{\mathrm{norm}}$ sets the interaction strength, and $\beta$ characterizes the nonlinear dependence on the Rydberg population. To account for a finite response time of the interaction-induced shift, the mean-field potential is taken to evolve dynamically according to
\begin{equation}
	\dot{V}_i^\text{MF} = 
	\frac{1}{\tau_\text{MF}}
	\left[
	V_{i}^\text{MF}(t + \Delta t)-V^\text{MF}_i(t)
	\right],
\end{equation}
over a timestep $\Delta t$ and where $\tau_\text{MF}$ is the mean-field response time. Because \(V_i^\text{MF}\) depends on the ensemble-averaged Rydberg population, all velocity classes are nonlinearly coupled through a shared, global feedback field. This collective feedback can partially synchronize the phases of different velocity groups, leading to coherent macroscopic oscillations of the system.

The Rabi frequency of the RF field, $\Omega_{RF}$ is
\begin{equation}
    \Omega_{RF} = \frac{\mathcal{p}}{\hbar}E_{RF} 
\end{equation}
where $\mathcal{p}$ is the RF transition atomic dipole moment and $E_{RF}$ is the electric field strength. For a heterodyned condition with an applied local oscillator electric field, $E_{LO}$, and a signal field $E_{SIG}$, the low frequency term is given by \cite{simons2019rydberg}
\begin{equation}
    E_{RF} = \sqrt{E_{LO}^2 + E_{SIG}^2 + 2E_{LO}E_{SIG}\cos{(\Delta}\omega t +\phi)}
\end{equation}
The dynamics of each velocity class are governed by a Lindblad master equation,
\begin{equation}
	\dot{\rho}^{(v)}(t)=
	-\frac{i}{\hbar}\big[H^{(v)}(t),\rho^{(v)}(t)\big]
	+\mathcal{L}\!\left[\rho^{(v)}(t)\right],
\end{equation}
with the dissipator
\begin{equation}
	\mathcal{L}[\rho]=
	\sum_{a>b}\gamma_{ab}
	\left(
	\sigma_{ba}\rho\sigma_{ab}
	-\frac{1}{2}\left\{\sigma_{aa},\rho\right\}
	\right),
\end{equation}
where \(\sigma_{ab}=|a\rangle\langle b|\) and \(\gamma_{ab}\) denotes the decay rate from state \(|a\rangle\) to \(|b\rangle\).

The resulting optical Bloch equations for the density-matrix elements of a given velocity class are~\cite{Wadenpfuhl:2023ywo, holloway2017electric},
\begin{align}
	\dot{\rho}_{11} &= \Omega_p\mathrm{Im}[\rho_{21}]
	+ \Gamma_{2}\rho_{22} \\[4pt]
	\dot{\rho}_{22} &= -\Omega_p\mathrm{Im}[\rho_{21}]
	+ \Omega_c\mathrm{Im}[\rho_{32}]
	- \Gamma_{2}\rho_{22}
	+ \Gamma_{3}\rho_{33}, \\[4pt]
	\dot{\rho}_{33} &= -\Omega_c\mathrm{Im}[\rho_{32}] + \Omega_{RF}\mathrm{Im}[\rho_{43}]
	- \Gamma_{3}\rho_{33} + \Gamma_4 \rho_{44}, \\[6pt]
	\dot{\rho}_{44} &= -\Omega_{RF}\mathrm{Im}[\rho_{43}]  - \Gamma_4 \rho_{44} \\[6pt]
	\dot{\rho}_{21} &= (i\Delta_p^{(v)} - \gamma_{21})\rho_{21} + \frac{i\Omega_p}{2}(\rho_{22} - \rho_{11}) - \frac{i\Omega_c}{2}\rho_{31}, \\[6pt]
	\dot{\rho}_{32} &=
	(i(\Delta_c^{(v)} - V_3^\text{MF}) - \gamma_{32})\rho_{32}
	+ \frac{i\Omega_c}{2}(\rho_{33} - \rho_{22}) \notag \\[-2pt]
	&\quad
	+ \frac{i\Omega_p}{2}\rho_{31} - \frac{i\Omega_{RF}}{2}\rho_{42},
\end{align}
\begin{align}
    \dot{\rho}_{31} &=
	(i(\Delta_p^{(v)} + \Delta_c^{(v)} - V_3^\text{MF}) - \gamma_{31})\rho_{32}
	+ \frac{i\Omega_p}{2}\rho_{32} \notag \\[-2pt]
	&\quad
	- \frac{i\Omega_c}{2}\rho_{21} - \frac{i\Omega_{RF}}{2}\rho_{41},\\[6pt]
    \dot{\rho}_{41} &=
	(i(\Delta_p^{(v)} + \Delta_c^{(v)} + \Delta_{RF}^{(v)} - V_3^\text{MF} - V_4^\text{MF}) - \gamma_{41})\rho_{41} \notag \\[0pt]
	&\quad
	+ \frac{i\Omega_p}{2}\rho_{42} - \frac{i\Omega_{RF}}{2}\rho_{31},\\[6pt]
    \dot{\rho}_{42} &=
	(i(\Delta_c^{(v)} + \Delta_{RF}^{(v)} - V_3^\text{MF} - V_4^\text{MF}) - \gamma_{42})\rho_{42}  \notag \\[0pt]
	&\quad
	+ \frac{i\Omega_p}{2}\rho_{41} + \frac{i\Omega_c}{2}\rho_{43} - \frac{i\Omega_{RF}}{2}\rho_{32},\\[6pt]
    \dot{\rho}_{43} &=
	(i(\Delta_{RF}^{(v)} - V_4^\text{MF}) - \gamma_{43})\rho_{43} + \frac{i\Omega_{RF}}{2}(\rho_{44}-\rho_{33}) \notag \\[0pt]
	&\quad
	+ \frac{i\Omega_c}{2}\rho_{42}
\end{align}
with $\rho_{ji} = \rho_{ij}^{\ast}$ and $\gamma_{ij} = (\Gamma_i + \Gamma_j)/2$.
\begin{figure}[h]
	\centering
	\includegraphics[width=\linewidth]{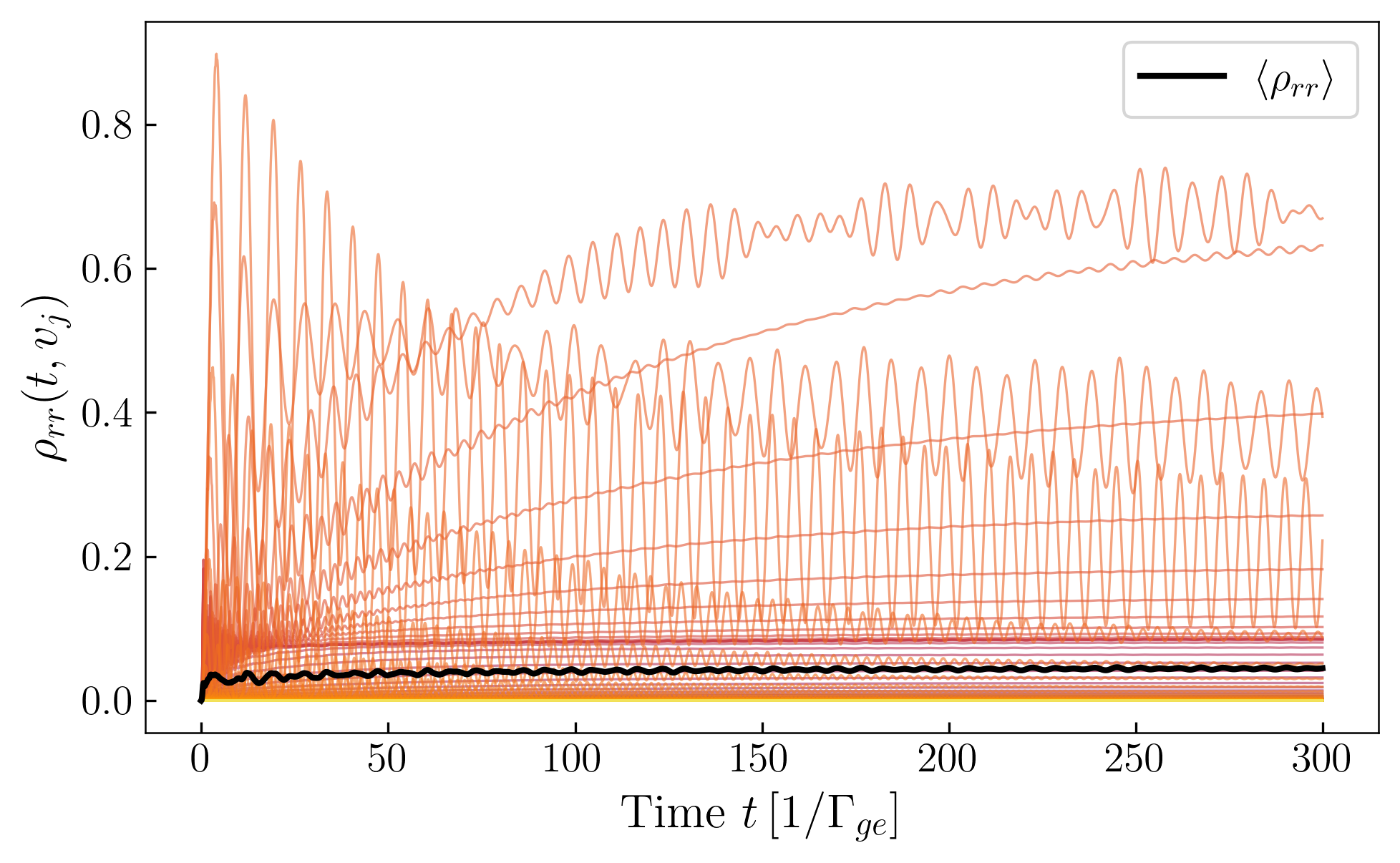}
	\caption{Velocity-resolved Rydberg state population as a function of time for $V_{\mathrm{norm}} = 400$, $\beta = 2.3$, $\Omega_p/\Gamma_{21} = 6~\mathrm{MHz}$, and $\Omega_c/\Gamma_{21} = 5~\mathrm{MHz}$.}
	\label{fig:theory_velocityresoloved}
\end{figure}
\begin{figure*}[htbp!]
    \includegraphics[width=\linewidth]{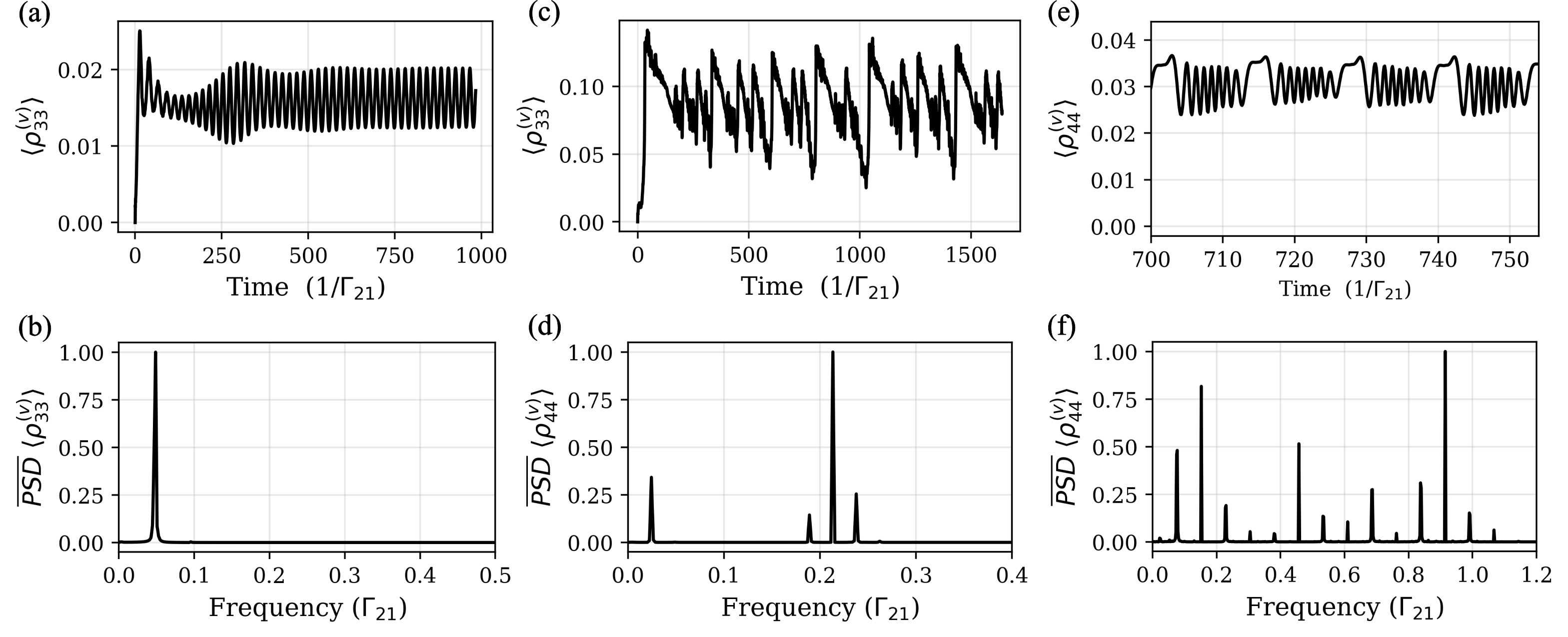}
    \caption{Numerical simulations from 4-level mean-field master equation model reproduce experimental observations. Throughout this image, we use the bracket notation $\langle \rho^{(v)}_{ii} \rangle$ to  refer to the average population over velocity classes and $\overline{PSD}$ to refer to the normalized power spectral density. (a-b) Stable oscillations without any applied RF. (c) Stable oscillatory regime without RF showing a triangle-like behavior where the Rydberg population slowly decays followed by an abrupt increase. (d) Intermodulation products for a heterodyned beatnote applied near an intrinsic oscillatory frequency.  (e-f) Frequency comb emergence from a heterodyned beatnote applied at an integer multiple of the intrinsic stable oscillation frequency. }
    \label{Fig:theory}
\end{figure*}
%
%This mean-field model does not quantitatively reproduce the experimental parameters and neglects additional atomic levels present in the experiment. Doppler averaging is included by explicitly evolving multiple velocity classes weighted by the Maxwell-Boltzmann distribution. The model nevertheless captures the essential nonlinear feedback mechanism responsible for the instability of the steady state and the emergence of stable self-sustained oscillations observed in the time-crystal regime.
%
%
Figure~\ref{fig:theory_velocityresoloved} shows the time evolution of the velocity-resolved Rydberg population \(\rho_{rr}(t,v_j)\) for different atomic velocity classes, together with the ensemble-averaged population \(\langle \rho_{rr} \rangle\). Individual velocity classes exhibit oscillations with different amplitudes and phases due to their distinct Doppler-shifted detunings, leading to dephasing at short times. However, because all velocity classes are coupled through the shared mean-field interaction shift $V_{i}^\text{MF}$, the dynamics are not independent. The global feedback provided by the population-dependent interaction shift partially synchronizes the velocity classes, suppressing complete dephasing and stabilizing a coherent macroscopic oscillation. As a result, while the velocity-resolved populations display large excursions and phase dispersion, the ensemble-averaged Rydberg population evolves into a stable, low-amplitude oscillatory state. This behavior illustrates how Doppler averaging and interaction-induced mean-field feedback together give rise to a robust self-sustained oscillation, consistent with the emergence of the time-crystal regime in the model.
\begin{table}[t]
\centering
\small
\begin{tabular}{lcccc}
\toprule
 & \multicolumn{4}{c}{Figure \ref{Fig:theory}} \\
\cmidrule(lr){2-5}
Parameter & (a-b) & (c) & (d) & (e-f) \\
\midrule
$\Omega_p$ (MHz)        & 200 & 150 & 300 & 200 \\
$\Omega_c$ (MHz)        & 80  & 125 & 360 & 80  \\
$\Omega_{LO}$ (MHz)     & 0   & 0   & 150 & 100 \\
$\Omega_{SIG}$ (MHz)    & 0   & 0   & 20  & 100 \\
$\Delta_p$ (MHz)        & -250 & -125 & -300 & -250 \\
$\Delta_c$ (MHz)        & -250 & -300 & -450 & -310 \\
$\Delta_{LO}$ (MHz)     & 0   & 0   & 0   & 0 \\
$\Delta\omega_{RF}$ (MHz) & 0 & 0   & 7   & 2.5 \\
$V_{\text{norm}}$ (GHz) & -20 & -20 & -16 & -20 \\
$\beta$                 & 1.8 & 1.5 & 2   & 1.8 \\
\bottomrule
\end{tabular}
\caption{Numerical simulation parameters used for Fig.~\ref{Fig:theory}.}
\label{table:numericalParams}
\end{table}
Our numerical simulations using the mean-field optical Bloch equations for a ladder system qualitatively reproduce many of the observed experimental dynamics (Fig. \ref{Fig:theory}), and in particular support the Rydberg-Rydberg mean-field interaction term given by equation \ref{V_MF}. The parameters used throughout Fig. \ref{Fig:theory} are given in Table \ref{table:numericalParams}. For no applied RF, figures \ref{Fig:theory}a-c show the time crystal phase characteristics changing from a nearly pure single toned oscillation frequency to a triangle-like waveform as observed experimentally in Fig. \ref{fig:1}b. Next, we consider an applied heterodyned RF. In Fig. \ref{Fig:theory}d, an the stable oscillatory frequency occurs at $\sim0.21\Gamma_{21}$ and a heterodyned beatnote is applied at $\sim0.24\Gamma_{21}$. Two intermodulation products are clearly seen at $\sim0.03\Gamma_{21}$ and $\sim0.18\Gamma_{21}$. Finally, as observed in Fig.\ref{Fig:theory}e-f, for specific combinations of the model parameters, the comb-like behavior manifests in the Rydberg population. In total, the mean-field model can reproduce many of the observed effects, but is highly dependent on the model parameters. Further work is needed to experimentally identify $V_\text{norm}$ and $\beta$, and in particular to understand if these parameters are constant or dependent on other Rabi rates and field detunings.

To gain further insight into the emergence of multiple frequency components and the comb-like spectra observed under RF heterodyne driving, we outline parallels between a classical analogue of a parametrically driven nonlinear oscillator.
\begin{figure}
	\centering
	\includegraphics[width=0.9\linewidth]{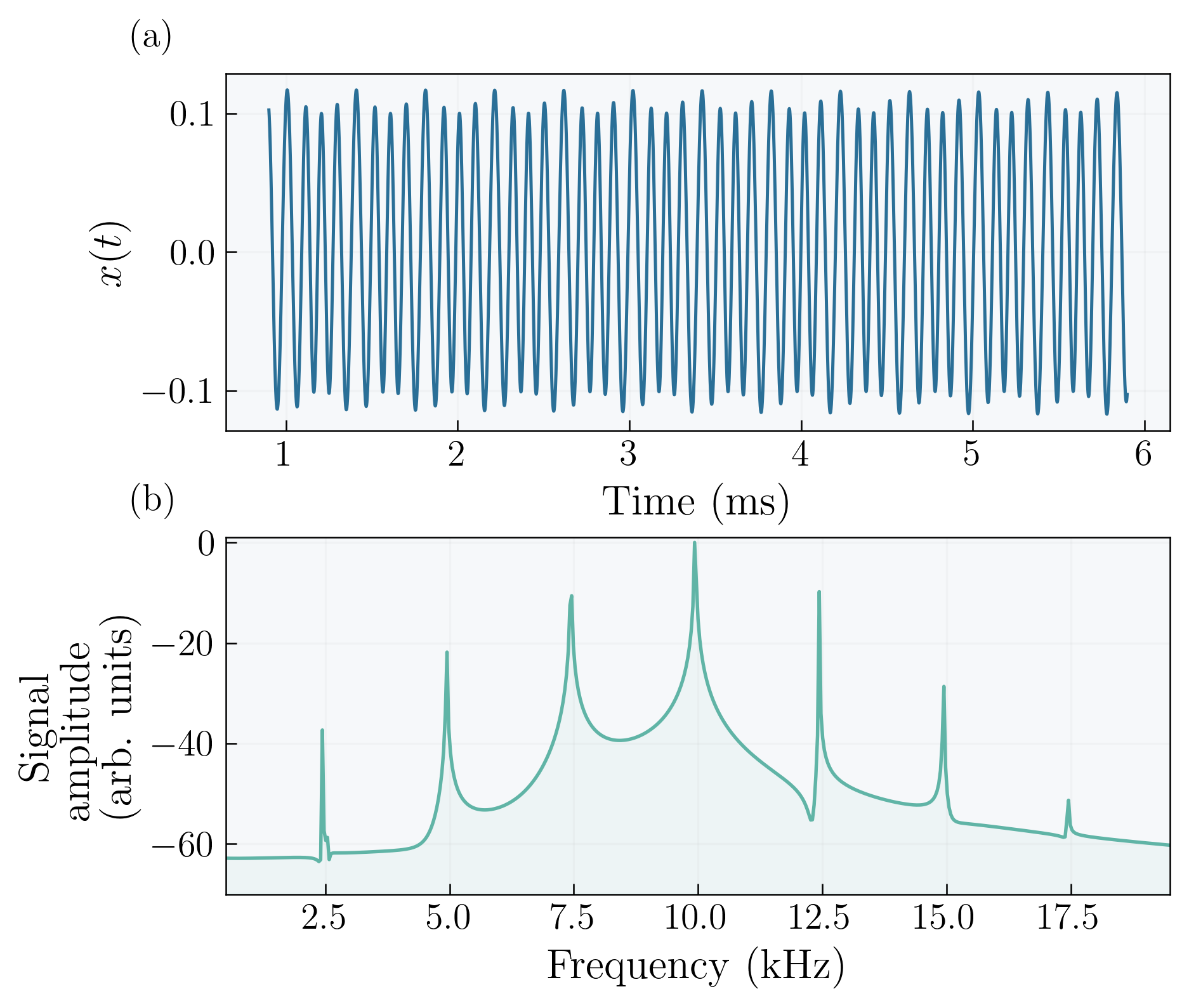}
	\caption{(a) Time-domain evolution $x(t)$ of a parametrically driven Van der Pol oscillator. (b) Fourier transform of the time-domain signal, showing the resulting frequency-domain comb structure.}
	\label{fig:classical}
\end{figure}
The equation of motion for a driven van der Pol oscillator is~\cite{strogatz1994nonlinear,Cook2019VanDerPol},
\begin{equation}
	\ddot{x}+\mu(x^2-1)\dot{x}+\omega_0^2\!\left[1+\epsilon\cos(\omega_d t)\right]x = 0,
\end{equation}
where \(\mu>0\) sets the nonlinear dissipation, \(\omega_0\) is the intrinsic oscillation frequency, \(\epsilon\) is the parametric driving strength, and \(\omega_d\) is the modulation frequency. Using experimental values, we set \(\omega_0 = \SI{10}{\kilo\hertz}\), \(\omega_d=\SI{2.5}{\kilo\hertz}\), and \(\epsilon=0.3\). The time-domain signal \(x(t)\) is shown in Fig.~\ref{fig:classical}(a), where the oscillator evolves into sustained oscillations in a stable limit-cycle regime. Here the nonlinear dissipation primarily stabilizes the oscillation amplitude, while the external modulation introduces additional frequency components.
The effect of the parametric modulation is to periodically perturb the limit-cycle dynamics, which generates sidebands around the main oscillation frequency at integer multiples of the modulation frequency. This produces a comb-like spectrum, as shown in Fig.~\ref{fig:classical}(b), with peak spacing set by \(\omega_d/2\pi\). In this sense, rf heterodyne driving in the experiment plays a role similar to an external periodic modulation of an intrinsic self-sustained oscillator, it can pull the oscillation frequency and generate multiple, regularly spaced spectral components through nonlinear mixing.
\\

%\section{Conclusions}
In summary, we demonstrate the emergence of a frequency comb in a driven-dissipative hot Rydberg cesium vapor under strong RF heterodyne driving in the time-crystal phase. RF driving enables tuning of the intrinsic oscillation frequency, frequency pulling, and the appearance of intermodulation products. In the absence of RF fields, the system exhibits multiple dynamical regimes, including bistability, Hopf bifurcations, and self-sustained oscillations. The frequency comb arises from parametric modulation of the intrinsic oscillations, producing a global feedback across all atomic velocity classes through mean-field interactions. These dynamics are captured by a four-level mean-field theoretical model and are further illustrated by a classical analogue using a driven Van der Pol oscillator. Our results establish interacting Rydberg ensembles as a tunable platform for exploring nonlinear many-body dynamics, synchronization, and frequency-comb generation for applications towards low-frequency electric field sensing and frequency stabilization.

\section*{Acknowledgments}
	This research was developed with funding from National institute of Standards and Technology. A contribution of the U.S. government, this work is not subject to copyright in the United States. 

\section*{Conflict of Interest}
The authors have no conflicts to disclose.

\section*{Data Availability Statement}
All of the data presented in this paper is available at https://doi.org/**/**

\bibliography{references}
\end{document}